\documentclass[graybox]{svmult}

\usepackage{makeidx}       % allows index generation
\usepackage{graphicx}       % standard LaTeX graphics tool
\makeindex             % used for the subject index
\usepackage{color,cite}

\def\nqq{\hspace*{-2em}}

\def\lal{&&\nqq {}}
\def\eq{Eq.\,}

\def\beq{\begin{equation}}
\def\eeq{\end{equation}}
\def\bear{\begin{eqnarray}}
\def\bearr{\begin{eqnarray} \lal}
\def\ear{\end{eqnarray}}
\def\earn{\nonumber \end{eqnarray}}

\def\nnn{\nonumber\\ \lal }

\def\qq{\qquad}

\def\e{{\,\rm e}}
\def\D{\partial}

\def\sign{\mathop{\rm sign}\nolimits}
\def\diag{\mathop{\rm diag}\nolimits}

\def\const{{\rm const}}
\def\eps{\varepsilon}

\def\then{\ \Rightarrow\ }

\newcommand{\vars}[1]{\left\{\begin{array}{ll}#1\end{array}\right.}

\def\rf#1{(\ref{#1})}
\def\mn{_{\mu\nu}}
\def\MN{^{\mu\nu}}
\def\mN{_\mu^\nu}

\def\M{{\bf M}}
\def\R{{\bf R}}

\def\oR{\overline{R}}
\def\og{\overline{g}}
\def\ME {\mbox{$\M_{\rm E}$}}
\def\MJ {\mbox{$\M_{\rm J}$}}
\def\GR{general relativity}

\def\ssph{static, spherically symmetric}

\def\wh{wormhole}
\def\whs{wormholes}
\def\asflat{asymptotically flat}
\def\emag{electromagnetic}

\begin{document}

\title*{Alexei Starobinsky and wormhole physics}
\author{Kirill A. Bronnikov\orcidID{0000-0001-9392-7558} and\\ 
		 Milena V. Skvortsova\orcidID{0000-0003-0812-3171}}
\institute{Kirill A. Bronnikov \at Rostest, Ozyornaya ul. 46, Moscow 119361, Russia;\\
		RUDN University, ul. Miklukho-Maklaya 6, Moscow 117198, Russia;\\
		National Research Nuclear University ``MEPhI'', 
		Kashirskoe sh. 31, Moscow 115409, Russia.
		\email{kb20@yandex.ru}
		\and Milena V. Skvortsova \at 
		RUDN University, ul. Miklukho-Maklaya 6, Moscow 117198, Russia.		
		\email{milenas577@mail.ru}}
\maketitle
% ==============================================

\abstract*{Alexei Starobinsky is most famous for his great contribution to cosmology, but
	he has considerable achievements in other branches of gravitational physics and 
	astrophysics, such as the theory of compact objects including black holes and wormholes.
	In this note, we give a brief review of Alexei's papers devoted to wormhole physics. 
	They mostly concern such issues of common interest as the necessary conditions for 
	wormhole existence in general relativity and its extensions as well as generic properties 
	of some kinds of wormholes.}

\vspace{-1cm}

\abstract{\\
	Alexei Starobinsky is most famous for his great contribution to cosmology, but
	he has considerable achievements in other branches of gravitational physics and 
	astrophysics, such as the theory of compact objects including black holes and wormholes.
	In this note, we give a brief review of Alexei's papers devoted to wormhole physics. 
	They mostly concern such issues of common interest as the necessary conditions for 
	wormhole existence in general relativity and its extensions as well as generic properties 
	of some kinds of wormholes. We also extend one of the no-go theorems on thin-shell
	wormholes to a wider choice of their symmetry and matter content.}

% ==================================
\section{Introduction} \label{sec:1}
% ==================================

  We would like to begin our paper with some memories of Alexei Alexandrovich 
  Starobinsky. He has been not only an outstanding scientist but a very attractive person,
  he paid very much attention to young scientists, students and postgraduates, when necessary,
  he helped them organizationally (for example, to take part in a school or conference of 
  interest). It was a typical practice for him to suggest a problem of interest to a young colleague 
  even if the latter worked in quite a different institute but showed a potential to solve it.      
  
  Alexei took part in a lot of conferences and was there very active, his talks did not only 
  contain the results of new research but always had an educational function, and he was always
  ready to answer questions, even if quite trivial for him. When he listened to talks by
  his colleagues, he usually sat in front of his computer, probably reading something there or
  answering messages, but when a talk was over, he used to ask questions and make comments 
  that were always appropriate and often deep.
  
  His interests were not restricted to cosmology, his main area: he had also outstanding 
  results in black hole physics (see, e.g, \cite{star5}), in quantum field theory in curved 
  space-time \cite{zeld}, and he expressed interest in many others areas. Among them 
  was \wh\ physics, a subject on which we had the honor and pleasure to be his co-authors. 
    
  Wormholes are nowadays quite a popular subject of both science fiction and serious 
  theoretical studies. These are hypothetic configurations of curved space-time having
  the shape of ``bridges'' between distant regions of the same universe or even between
  different universes. The possible existence of such objects in our Universe would lead
  to diverse remarkable effects, beginning with observation of the sky of another universe
  through a \wh\ throat and ending with such a fantastic construct as a time machine.  
  It is therefore not too surprising that, although nobody has ever confidently observed 
  any \wh, many researchers study and discuss both the conditions of \wh\ emergence 
  and existence and their possible observational signatures.
  
  As already said, Alexei Starobinsky expressed active interest in \wh\ physics. His papers in 
  this area, written together with the present authors \cite{star1,star2,star3}, are as a whole 
  not devoted to obtaining or discussion of particular \wh\ solutions of different theories of 
  gravity (whose list is now extremely rich) but to such questions of general interest as the 
  \wh\ existence conditions and their main properties that do not depend on a particular 
  model. The main purpose of the present paper is to recall the main results of those papers, 
  emphasizing Alexei's contribution to this intriguing sector of gravitational physics. For one
  of the no-go theorems of \cite{star1,star2,star3}, that on \ssph\ thin-shell wormholes,
  we present an extension to wider symmetry properties and arbitrary matter sources.  
  
% =================================================  
\section{Throats and asympotic regions}\label{sec:2}
% =================================================

  It is quite well known that the existence of static \wh\ throats (2-surfaces of minimum area
  in these space-time tunnels) in \GR\ inevitably requires a material source of gravity whose 
  stress-energy tensor (SET) $T\mN$ violates the weak and null energy conditions (WEC and 
  NEC) \cite{MS88,HV97}, and such sources have acquired the name of ``exotic matter,''
  also often called phantom or ghost matter (we use the words `ghost' and `phantom' as
  synonyms). There are strong reasons to suppose that Nature, 
  if not  completely forbids the existence of such matter, still tries to avoid its emergence. 
  Therefore, much effort is devoted to a search for \wh\ models without such matter.     
    
  One can point out two ways in attempts to achieve this goal: either consider nonstatic 
  models in \GR\  (for example, in the framework of simple cosmological models), or use 
  theories of gravity other than \GR. Both ways are widely used and discussed by the
  researchers. The second way certainly implies that the corresponding theory alternative
  to \GR\ is sufficiently well founded and supported by observations.    
  
  Scalar-tensor theories (STT) of gravity are among the most promising alternative theories,
  and their numerous models sufficiently well agree with observations (see \cite{stt-rev1,stt-rev2} 
  for recent reviews). By the year 2005, when we began our work \cite{star1}, the analysis 
  of current cosmological observations revealed some evidence that the Dark Energy (DE), 
  responsible for the acceleration expansion of the Universe if the Einstein equations are 
  assumed to hold, may be well described by a phantom matter equation of state parameter
  $w \equiv p_{\rm DE}/\rho_{\rm DE}$, slightly smaller than $-1$. (Let us note that this estimate
  remains relevant today, even though models with a cosmological constant $\Lambda$ 
  ($w=-1$) remain most popular and comprise the `standard' ($\Lambda$CDM) cosmological
  paradigm.) It was, however, clear that the possibly observed $w < -1$ does not necessarily
  mean the existence of ghosts in Nature since there are STT models without ghosts that admit 
  a phantom DE behavior and even a super-acceleration of the present Universe, such that 
  $dH/dt > 0$, where $H(t)$ is the Hubble parameter in Friedmann-Robertson-Walker 
  cosmological models. For example, this may take place in the framework of the  
  Brans-Dicke STT with a nonzero scalar field potential. Thus, as was shown explicitly 
  \cite{star-BD}, such a model possesses sufficient freedom to account for all observations 
  justifying the presently possible phantom-driven expansion as well as those concerning 
  the luminosity distances and inhomogeneity growth in the Universe.
  
  So, if a nonphantom scalar field in STT can provide a phantom behavior in cosmology, a 
  natural question is: can such a field create local entities also usually associated with phantoms, 
  i.e., wormholes? Our paper \cite{star1} gives a negative answer to this question, thus 
  extending some previous results \cite{vac5, BG05},
  
  To begin the discussion, it is necessary to define its subject.
  We here, following \cite{star1, star2, star3}, consider static space-times and say that that the 
  geometry is that of a {\bf \wh}\ if (i) the space-time is globally regular and free from horizons, 
  (ii) its spatial sections contain a 2D surface of minimum area, called a {\bf throat}, and (iii) 
  the throat is a member of a family of 2D surfaces whose area is not bounded above on 
  both sides from the throat. In the simplest case of spherical symmetry this means that a throat
  is a sphere of minimum spherical radius $r$, and that $r \to \infty$ on both sides far from this 
  sphere. Asymptotic flatness on one or both `ends' is not necessary though often desirable.
  
  This definition is more narrow than those used by a number of authors but covers the cases 
  of interest in many studies.  
  
  Now let us briefly recall the main points of our reasoning in \cite{star1}.  
  The main feature of general (Bergman-Wagoner-Nordtvedt) STT 
  is a nonminimal coupling between a scalar field $\Phi$ and the curvature. 
  The Lagrangian in a (physically preferred) Jordan-frame manifold \MJ\ with
  the metric $g\mn$ may be written as\footnote
  		{We use the following conventions: the metric signature $(+{}-{}-{}-)$; 
  		the  Riemann and Ricci tensors defined as
		$R^{\sigma}{}_{\mu\rho\nu} = \D_\nu\Gamma^{\sigma}_{\mu\rho}-\ldots$ and
	      $R\mn = R^{\sigma}{}_{\mu\sigma\nu}$, so that the Ricci scalar
	      $R = g\MN R\mn$ is positive for both de Sitter space-time and matter-dominated
		 cosmology; the system of units $8\pi G = c = 1$.}
\beq
	2L= f(\Phi) R + h(\Phi)g^{\mu\nu}\Phi_{,\mu} \Phi_{,\nu}
				- 2U(\Phi) - F\MN F\mn,   						\label{Lagr}
\eeq
  where $R$ is the Ricci scalar and $F\mn$ the electromagnetic field tensor;
  the functions $f$, $h$ and $U$ are arbitrary. The fermion matter Lagrangian 
  is assumed to have no coupling with $\Phi$, hence fermion masses are constant,
  and atomic clocks measure the proper time $t$ in \MJ. It is, however, useful 
  to also consider the Einstein frame, defined as a manifold \ME\ with the metric
\beq
	\og\mn = |f(\Phi)| g\mn,                             \label{conf}
\eeq
  in which the Lagrangian \rf{Lagr} takes the form
\beq
        2L_E = (\sign f) \bigl[\oR
                    + (\sign \ell) \og\MN \phi_{,\mu}\phi_{,\nu}\bigr]
		    	    - 2V(\phi) - F\MN F\mn, 		\label{LE}
\eeq
  where overbars mark quantities obtained from or with $\og\mn$, indices are raised 
  and lowered with $\og\MN$ and $\og\mn$, and the functions involved are related by
\beq
    	\ell(\Phi) := fh + \frac{3}{2}\biggl(\frac{df}{d\Phi}\biggr)^2,
    \qq
    	\frac{d\phi}{d\Phi} = \frac{\sqrt{|\ell(\Phi)|}}{f(\Phi)},
    \qq
	V(\phi) = |f|^{-2} U(\Phi).                          \label{trans}
\eeq

  Let us require that the theory has no ghosts, hence $f(\Phi) > 0$ 
  (the graviton is not a ghost) and $\ell(\Phi) > 0$ (the $\Phi$ field is not a ghost).
  Moreover, let us assume that $f(\Phi)$ and $\ell(\phi)$ are smooth and
  bounded above and below by positive constants. According to \cite{HV97}, 
  a static \wh\ throat (defined as a minimal 2D surface in a 3D manifold) necessarily 
  requires that the matter source the Einstein equations violates the NEC
  that requires $T\mn \xi^\mu \xi^\nu \geq 0$, where $T\mn$ is the SET of matter, 
  and $\xi^\mu$ is an arbitrary null vector.  Since the matter sources in (\ref{LE}) satisfy 
  the NEC, we can assert that \whs\ (and even \wh\ throats) are impossible in \ME.
  
  Now we notice that under our assumptions, the conformal mapping 
  $\og\mn = f(\Phi)g\mn$ transfers a flat spatial infinity in one frame to a flat spatial 
  infinity in the other. Therefore, if there is a \wh\ in \MJ, which is \asflat\ on each
  side of the throat, then its each flat infinity maps to its counterpart in \ME, the 
  whole manifold is smooth, and we thus obtain a \wh\ in \ME, contrary to the
  above-stated impossibility. This contradiction makes us conclude that static 
  \asflat\ \whs\ are absent in the Jordan frame as well.

  In this simple reasoning, no spatial symmetry assumption was made, hence any 
  static \whs\ are ruled out in the theory (\ref{Lagr}), if everywhere $f > 0$ and $\ell > 0$. 
  Moreover, no assumption was made on $U(\Phi)$ (apart  from smoothness 
  and compatibility with asymptotic flatness). So the result is quite general. 
  
  The no-\wh\ statement for \MJ\ can be further generalized in a few respects \cite{star1}:
\begin{description}[\hspace{5mm}]  
\item[(i)]  
	our inferences remain valid in the presence of any other ghost-free (respecting the NEC)
	matter; 
\item[(ii)] 
	instead of asymptotic flatness, we can simply require the existence of two spatial infinities 
	(defined in terms of infinitely growing areas of closed 2-surfaces in the spatial sections of 
	\ME\ and \MJ); 
\item[(iii)] 
	the no-\wh\ statement holds not only in STT but in any metric theory of gravity 
	(e.g., $F(R)$ or $F(R, \Phi)$ \cite{F(R), maeda}) admitting a conformal mapping of the 
	physical manifold \MJ\ to \ME\ where GR holds with ghost-free matter, 
	provided that the conformal factor is everywhere smooth and positive;  
\item[(iv)] 
	under certain restrictions, our statement can even be extended to dynamic \whs\ 
	based on the no-\wh\ results for \ME\ obtained by Hochberg and Visser in \cite{HV98}. 
\end{description}
  
  The latter results rest on the definition of a dynamic wormhole throat as a marginally 
  anti-trapped surface. There are other definitions (see, e.g, \cite{dyn-def} for a review), 
  some of them are less restrictive than that of \cite{HV98}. Thus, a number of dynamic 
  \wh\ solutions in GR have been obtained using an intuitively clear definition of a throat 
  referring to a physically preferable set of constant-time spacelike surfaces, see, e.g.,
  solutions describing \whs\ in a cosmological background in the framework of GR, sourced 
  by such ghost-free kinds of matter as nonlinear electrodynamics \cite{NED-lobo, NED-kb} 
  and dust \cite{BKS23}, but none of such objects can exist eternally and have initial and/or
  final singularities. 
  
  Returning to static systems in (not only) STT, we should stress that throats are not ruled 
  out in \MJ\ (see an example below in this paper): it is a \wh\ as a global entity that is ruled 
  out. Therefore, nothing prevents one from constructing, for instance, composite \wh\ 
  models where a neighborhood of the throat is described by an appropriate STT solution, 
  while the rest of space is either empty or has some other ghost-free matter as a source, 
  admitting asymptotic flatness.

  A further discussion in \cite{star1} concerned possible weakening of the assumptions on the 
  function $f(\Phi)$. It is shown that if $f(\Phi)$ somewhere reaches zero, there can be exceptional 
  twice \asflat\ \wh\ solutions in some STT, conformally mapped to solutions with extremal 
  horizons in \ME, but such explicit examples are yet to be found. Furthermore, if $f(\Phi)$ is 
  allowed to become negative (the effective gravitational constant is negative in such regions), 
  then \wh\ solutions are not ruled out, and a few examples of such solutions are known
  \cite{kb73, barc00, step02}, with sources in the form of nonminimally coupled massless 
  scalar fields and possible inclusion of an \emag\ field, such that in \rf{Lagr} we have 
  $f(\Phi) = 1 - \xi \Phi^2$ ($\xi = \const$), $h(\Phi) \equiv 1, \ U(\Phi) \equiv 0$.  
  All such solutions involve the phenomenon of conformal continuation \cite{kb02}. The 
  point is that the surface where $f =0$ is a singularity in \ME\ but regular in \MJ, and the
  conformal mapping \rf{conf} from the whole \ME\ leads to only a part of \MJ, which 
  therefore has to be continued to a new region with another sign of $f$.  Apart from 
  a special nature of such constructs with respect to the whole set of solutions of a particular 
  theory, it also happens that such wormhole solutions are generically unstable under 
  spherically symmetric (monopole) perturbations \cite{BG05, step02}. 
  
  Thus we have the following {\bf no-go theorem} \cite{star1}: in any STT (or $F(R)$, or 
  $F(R, \Phi)$ theories of gravity ), in any systems free of ghosts, no complete, infinitely 
  extended static \wh\ solutions are allowed in the physical Jordan conformal frame \MJ, 
  even though throats, in general, can be present. 
  
  If the no-ghost conditions are weakened as described above, then special \wh\ solutions 
  can exist, but none of them can be regarded realistic. 
  
  It remains to add that if the scalar $\Phi$ is phantom, that is, if in \rf{trans} $\ell(\Phi) < 0$,
  \wh\ solutions in STT are readily obtained, as confirmed by numerous examples beginning 
  with \cite{kb73}.     

% ======================================
\section{Thin shells and wormholes}
% ======================================

  After the publication of \cite{star1}, attempts to obtain static phantom-free \wh\ models 
  continued. In particular, it was claimed that at least in some cases, if a \wh\ throat is
  provided by a thin shell of matter (e.g., in the Brans-Dicke STT in a certain range of 
  the coupling constant $\omega$), the model may satisfy the NEC and WEC \cite{eir08}.
  Our paper \cite{star2}, aimed at making clear this issue, was a kind of response to such 
  claims. As was explicitly shown in \cite{star2}, in {\it any\/} STT with a massless 
  non-ghost scalar field, in {\it any\/} thin-shell \wh\ constructed from two identical 
  regions of vacuum \ssph\ space-times, the shell that makes the throat has a negative
  surface energy density. We here extend this result, remaining in the framework of static 
  spherical symmetry but canceling the requirement of vacuum ($L_m =0$), 
  that of $U(\Phi)=0$, and that of identity of the two regions.
  
  The proof in \cite {star2} again uses the mapping \rf{trans} and the fact that in the Einstein 
  frame \ME, trying to build a thin-shell \wh, we inevitably obtain a negative surface density
  of the shell. Under the assumptions made in \cite{star2} (a massless non-ghost scalar 
  field and no other matter), in \ME\ we are dealing with Fisher's scalar-vacuum solution
  \cite{fish} with the metric 
\bearr                \label{ds-fish}
		ds^2 = P^a dt^2 - P^{-a} dx^2 - P^{1-a}r^2 d\Omega^2,
\nnn		
      P = P(x) := 1 - \frac{2k}{x},\qq   d\Omega^2 = d\theta^2 + \sin^2\theta d\varphi^2,
\ear
  where $a$ and $k > 0$ are constants, $|a| < 1$, and $x$ is the so-called quasiglobal 
  (or Buchdahl) radial coordinate chosen so that $g_{tt} g_{xx} = -1$. The range of $x$ is 
  $2k < x <  \infty)$, and $x=2k$ is a central (since the spherical radius 
  $r = \sqrt{-g_{\theta\theta}}$ has there a zero value) naked singularity. 
  
  A thin-shell \wh\ is obtained from \rf{ds-fish} as follows: (i) consider a region (A) 
  of the space-time \rf{ds-fish} where $x > x_0$, and $x_0 = \const > 2k$, so that 
  $r(x_0) > 0$; (ii) consider one more copy (B) of the same region; (iii) introduce the new 
  coordinate $y$ such that $y = x - x_0$ in region (A) and $y = x_0 -x$ in region (B); 
  (iv) unify the two regions by identifying their spheres $y =0$. Then it is directly verified 
  by a substitution to the Einstein equations that the obtained geometry corresponds 
  to the SET in which the energy density component has the form
\beq                       \label{E-shell}
			T^t_t =  -\frac{4\delta(y)}{x_0^2} 
					\bigg(1 - \frac{2k}{x_0}\bigg)^{a-1} [x_0 - (1+a)k] 
					+ \mbox{a finite quantity},
\eeq   
  where $\delta(y)$ is Dirac's delta function. This clearly means that the junction 
  surface $y =0$ has a negative surface density. The same result is obtained if we 
  construct a thin-shell \wh\ using the electrically or magnetically charged extension 
  of Fisher's space-time obtained by Penney \cite{penney}.
  
  Suppose that we have a \ssph\ thin-shell \wh\ solution in an STT \rf{Lagr} with $U = L_m =0$.
  Then, after the mapping \rf{trans}, we obtain the (electro)vacuum problem in GR, 
  on each side of the throat, hence these both regions will be desccribed in \ME\ by the
  Fisher metric or its charged extension due to Penney. Now, it is necessary to take into
  consideration that the stress-energy tensor $T\mN$ of any kind of matter (other than the
  nonminimally coupled scalar $\Phi$) present in \MJ\ transforms into \ME\ by the simple law  
\beq           \label{trans-T}
			T\mN{}^{(J)} = f^2(\Phi) T\mN{}^{(E)},
\eeq  
  where the superscripts $(J)$ and $(E)$ mark the SETs in the corresponding conformal frames. 
  In particular, this law is valid for the surface SET of the thin shell.
  
  Assuming, as before, that $f(\Phi)$ is everywhere finite and positive, we come to the
  conclusion that due to \eq \rf{E-shell} (where $T\mN = T\mN{}^{(E)}$), the surface energy
  density of the shell is also negative in \MJ. Thus we obtain that, in fact, the no-go theorem 
  of Section 2 is valid as well for thin-shell \whs, though a proof \cite{star2} has been obtained 
  under rather strong restrictions (static spherical symmetry, vacuum or electrovacuum, and 
  identity of regions on different sides of the throat).
  
  Let us now try to weaken these restrictions: assuming, as before, static spherical symmetry, 
  let us only suppose that we are working in GR (or, equivalently, in \ME\ of some STT) with 
  matter respecting the NEC, and use the corresponding metric to obtain a thin-shell \wh\ by 
  the cut-and-paste method as described above.   
  
  To obtain the proof, let us notice that for the general \ssph\ metric written in terms of the 
  quasiglobal coordinate $x$,
\beq
		ds^2 = A(x) dt^2 - \frac {dx^2}{A(x)} - r^2(x) d\Omega^2,
\eeq  
  the difference of the $({}^t_t)$ and $({}^x_x)$ components of the Einstein equations reads
\beq                 \label{01}
		2 A \frac{r''}{r} = - (\rho + p_r),
\eeq  
  where the prime means $d/dx$, $\rho$ and $p_r$ are the energy density and radial 
  pressure of matter, respectively, and we have $A(x) >0$ and $r(x) > 0$ at regular points. 
  The NEC component with the radial null vector $\xi^\mu$ 
  requires $\rho + p_r \geq 0$, therefore, \eq \rf{01} leads to $r'' \leq 0$ in any regular static 
  region with non-phantom matter. 
  
  Now suppose that there is a thin-shell \wh\ with the throat at $x=0$ (without loss of 
  generality), where $A$ and $r$ are finite, and two spatial infinities ($r\to \infty$) at 
  $x \to \pm \infty$. Then, since $r'' \leq 0$, at negative $x$ the function $r(x)$ decreases 
  from infinity with an everywhere negative derivative $r'(x)$, while at $x>0$ there is
   everywhere $r'(x) > 0$. It means that at $x =0$, where they meet, we have inevitably a 
   finite jump of  $r'(x)$ from a negative value to a positive one, which implies that 
   $r''(x) \propto \delta(x)$ with 
   some positive proportionality factor. Then, according to \rf{01}, the quantity $\rho + p_r$    
   is a multiple of $\delta(x)$ with a negative factor. However, the whole SET of a thin shell
   of fixed radius $r$ has the form 
\beq   
   	   T\mN{}^{\rm (shell)} = N\delta(x) \diag (\sigma,\ 0,\ -p_\bot,\ -p_\bot), 
\eeq   
   where $\sigma$ is the surface density, $p_\bot$ is the tangential pressure of the shell 
   matter, and $N>0$ is a factor depending on the particular values of $A(0)$ and $r(0)$. 
   The radial  pressure is zero since the shell is not extended in the radial direction. 
   From $\rho + p_r < 0$ we conclude that the shell has a negative surface density, $\sigma < 0$.
   
   At this point we can actually repeat the reasoning used in Section 2: if the corresponding 
   solution of any metric theory of gravity (STT or any other) whose metric is connected 
   with ours by an everywhere finite and positive conformal factor $f$, due to \rf{trans-T} 
   there will be again a thin shell with negative surface density, violating the WEC and NEC. 
   We conclude that in all such theories it is impossible to construct a \ssph\  thin-shell \wh\ 
   without NEC and WEC violation.  
   
   A curious feature of our proof is that we did not apply the well-known Darmois-Israel 
   formalism for describing thin shells dynanics: the present method is simpler and quite 
   sufficient for  static shells and is rather similar to the method applied in \cite{star-sok} 
   to the case of cosmic strings described by conical singularities.
   
   This proof did not suppose any particular form of matter filling the space-time, only assuming 
   its non-phantom nature. Moreover, the matter and the metric can be quite different `to the left'
   and `to the right' of the throat. Thus we have significantly strengthened the no-go theorem 
   under consideration for thin-shell \whs. 
  
% ========================================
\section{More observations on STT and $F(R)$ theories}
% ========================================
   
   Our third paper \cite{star3} was also first motivated by some wrong or doubtful claims met in 
   the literature that required making clear some subtle points of \wh\ theory. As often happens 
   in such cases, the study has led to some more general observations of interest.  
  
   Such statements mostly concerned the Brans-Dicke (BD) STT \cite{BD} that corresponds 
   to the choice 
\beq                   \label{BD}
		f(\Phi) = \Phi, \qq  h(\Phi) = \frac{\omega}{\Phi}
\eeq   
  in the Lagrangian \rf{Lagr}, where $\omega$ is a coupling constant. According to \rf{trans},
  the $\Phi$ field is phantom if $\omega < -3/2$, it is canonical if $\omega > -3/2$, and it 
  has no dynamics if $\omega = -3/2$. Thus, as is clear, e.g., already from \cite{kb73}, vacuum 
  static \wh\ solutions of BD theory exist for all $\omega < -3/2$ and can only emerge in 
  exceptional cases if $\omega > -3/2$. Meanwhile,  
\begin{itemize}
\item
    In \cite{BD1} it was asserted that vacuum wormhole solutions can be found in the BD theory 
    at $\omega$ belonging to the non-ghost range $(-3/2,-4/3)$.
\item    
    A similar conclusion was made in \cite{BD2} for a wider non-ghost range of $\omega$,
    including $\omega =0$.
\item    
    In \cite{BD1}, it was also asserted that vacuum BD wormholes are possible for
    $\omega < -2$ without mentioning the range $(-2, -3/2)$.
\item    
    In \cite{BD3}, wormhole solutions were indicated for some $F(R)$ theories, which are 
    known to be equivalent to the BD theory with $\omega=0$ and a nonzero potential
    $U(\Phi)$. 
\end{itemize}

   Apart from general objections such as that some authors use (unlike us) the term `\wh' for 
   any space-time containing a throat, the above-mentioned statements of \cite{BD1, BD2, BD3}
   required a closer analysis of BD vacuum solutions and a comparison of their different
   parametrizations. Let us reproduce the main points. 
   
   First of all, recalling \cite{kb73}, we can write the general vacuum \ssph\ solution 
   of any STT \rf{Lagr} with $U \equiv 0$ as follows:
\bearr              \label{sol-J}
		ds^2_J = \frac 1 {f(\Phi)} \bigg\{\e^{-2mu} dt^2 - \frac{\e^{2mu}}{s^2(k,u)}
                    			\bigg[\frac{du^2}{s^2(k,u)} + d\Omega^2\bigg]\bigg\},
\nnn            
		 \phi = Cu, \qq   s(k,u) := \vars     {
                            k^{-1}\sinh ku,  \ & k > 0, \\
                                 u,          \ & k = 0, \\
                            k^{-1}\sin ku,   \ & k < 0,    }                        			
\ear     
   where the fields $\Phi$ (involved in the Lagrangian \rf{Lagr} written for \MJ) and $\phi$
   (involved in \rf{LE} written for \ME) are related as indicated in \rf{trans}.
   The index $J$ points out that it is the Jordan-frame metric. The integration
   constants $C$ (scalar charge), $m$ and $k$ are connected by the relation 
\beq                                                           \label{int}
        2k^2 \sign k = 2m^2 + \eps C^2, \qq \eps = \sign \ell.
\eeq   
   The coordinate $u$ used in \rf{sol-J} is the harmonic radial coordinate in \ME,
   defined by the coordinate condition $\og_{uu} = \og_{tt} \og^2_{\theta\theta}$, and the
   expression in the curly brackets is nothing else than the metric in \ME. 
   
   The branch $k < 0$ of the solution (existing only in the phantom case $\eps =-1$) 
   describes \whs\ as long as $f(\Phi)$ is everywhere positive and finite since two adjacent 
   zeros of the function $\sin ku$ (e.g., $u=0$ and $u = \pi/|k|$) describe flat spatial 
   infinities, and this fact is not spoiled by the regular conformal factor $1/f(\Phi)$. This is true, 
   in particular, for the BD solution in which $\eps = \sign (\omega + 3/2)$ and
 \beq                             \label{Phi}
      	f(\Phi) = \Phi = \Phi_0\exp (\phi/\sqrt{|\omega+3/2|}),  \qq
           \Phi_0 = \const,    
\eeq
    is finite at any finite $\phi = Cu$. This removes any doubt concerning the \wh\ existence 
    at any $\omega < -3/2$. A more detailed description of \ssph\ BD solutions 
    (not only \wh\ ones) with $\omega < -3/2$ may be found in \cite{cold1, cold2}.
    
   Let us now consider the canonical case $\eps = +1$, $\omega > -3/2$. In this case, the 
   expression in curly brackets in \rf{sol-J}, in which $k>0$, is Fisher's metric written 
   using the harmonic coordinate $u$, as verified by the substitution 
\beq
		\e^{-2ku} = 1 - \frac{2k}{x} \equiv P(x), \qq a = \frac mk < 1,
\eeq      
  which leads the metric in \ME\ to the form \rf{ds-fish}. For the BD scalar field we then 
  obtain 
\beq
		\Phi = P^\xi, \qq   \xi = -C/\big(2k\sqrt{\omega + 3/2}\big),
\eeq    
   and the relation (\ref{int}) turns into
\beq                                                            \label{int+}
       (2\omega + 3) \xi^2 = 1 - a^2.
\eeq  
   The BD metric in \MJ, defined and regular at $x > 2k >0$, has the form  
\beq             \label{ds-BD}
   ds^2_J = P^{-\xi} \Big[ P^a dt^2 - P^{-a} dx^2 - P^{1-a} x^2 d\Omega^2 \Big].   
\eeq  

  To answer the question, whether or not this metric can describe a \wh, one should find 
  out (1) whether there can be a minimum of the radius $r_J = x P^{(1-a-\xi)/2}$, and 
  (ii) if yes, is such a manifold globally regular and have an infinite radius at both ends.
  Since the metric \rf{ds-BD} is \asflat\ at $x \to \infty$, it remains to study its behavior 
  as $x \to 2k$.  
  
  The first question has been answered as follows \cite{star2}: throats certainly do not 
  exist in all cases, but still appear at arbitrarily large values of the parameter $\omega$
  (even though the limit $\omega \to \infty$ corresponds to GR). To prove that, let us single
  out the subfamily of solutions given by $\xi = 2(1-a)$. Then it is easy to obtain that 
  $r_J(x)$ has a minimum at $x = k(3-a) > 2k$ (because $a <1 \then 3-a >2$). Also, 
  from (\ref{int+}) we find
\beq                        \label{BD-thr}
		     2\omega + 3 = \frac{1+a}{4(1-a)},
\eeq
  and taking $a$ close enough to unity, the constant $\omega$ can be made arbitrarily large.
  
  The metric behavior at $x=2k$ is rather diverse but is generically singular in some sense,
  see details in \cite{star2}. The only exception, making the sphere $x=2k$ regular, is the 
  case $a = \xi =1/2 \ \then \ \omega=0$. The metric (\ref{ds-BD}) then 
  has the so-called `spatial Schwarzschild' form
\beq                                                      \label{spa-S}
     ds_J^2 = dt^2 - \frac{dx^2}{1 - 2k/x} - x^2 d\Omega^2
            \equiv dt^2 - 4(2k + y^2)dy^2 - (2k + y^2)^2 d\Omega^2,
\eeq
  where $y^2 = x - 2k$. The range of $y$ is $y \in \R$, so that the manifold \MJ\ consists 
  of two copies ($y >0$ and $y < 0$) of a region mapped from \ME\ defined for $x > 2k$,
  joined by the regular sphere $y=0$, the throat.  It is a \wh\ obtained owing to the 
  conformal continuation phenomenon \cite{kb73, kb02}. For the scalar field  we have 
  $\Phi = y/\sqrt{y^2 + 2k}$, it is zero at the transition sphere $y=0$ and negative beyond it,  
  which means that the effective gravitational constant is negative there.

  Thus in BD theory we have a full confirmation of the results of \cite{star1} on \whs\
  in STT. As to $F(R)$ theories of gravity, the results obtained for STT are easily 
  reformulated for them using the well-known correspondence between these two classes
  of theories. Specifically, if we write the Lagrangian of $F(R)$ theory as 
  $L = F(R)/2 + L_m$, then, assuming
\beq
		f(\Phi) = \Phi = F_R, \qq h(\Phi) =0, \qq 2U(\Phi) = R F_R -F
\eeq      
  (where $F_R \equiv dF/dR$),
  we obtain an equivalent representation of the same theory as the BD theory with the 
  coupling constant $\omega =0$, where the parametrization of the $\Phi$ field is chosen 
  as $f(\Phi) \equiv \Phi$. Note that these both form of the theory are presented in the 
  same Jordan frame \MJ, with the same metric $g\mn$, and they have a common Einstein 
  frame \ME. The condition $f >0$ in STT transforms into $F_R > 0$; furthermore, the function 
  $\ell (\Phi)$ in \rf{trans} is nonnegative but can become zero at special values of $R$,
  where $F_{RR} \equiv d^2F/dR^2 =0$; at such points, the parametrization $f = \Phi$ 
  is lost. Moreover, translating the results of \cite{star1} (and Section 2 in the present paper)
  to the `language' of $F(R)$ theory, we can assert that \wh\ solutions are only possible if 
  $F(R)$ has an extremum at some $R=R_0$ and may be obtained by a conformal continuation 
  through such a surface, which is regular in \MJ\ but singular in \ME\ \cite{kb-ch05}. Beyond     
  such a surface there is a region where $\Phi = F_R <0$, with a negative effective 
  gravitational constant. 
  
  There is a subtle point \cite{star3} if at some value of $R$ there is $F_{RR}=0$ while 
  $F_R \ne 0$: such cases are generically singular, with finite algebraic curvature invariants 
  but diverging differential invariants. A similar situation in cosmological solutions is discussed 
  in \cite{star6}. In exceptional cases such surfaces can be regular, but global \wh\ solutions  
  are still impossible by the no-go theorem of \cite{star1}.
  
  What about examples of \wh\ solutions in $F(R)$ theory that can be found in 
  the literature? In most of them, only the existence of throats is considered, and, as discussed 
  above (see especially \eq \rf {BD-thr}), they are not at all excluded even in ghost-free cases.
  Ref. \cite{BD3} discusses some examples of \wh\ solutions of $F(R)$ theory in the presence 
  of matter with anisotropic pressure, but in some of these examples there is $F_R <0$
  in certain regions, while others contain singularities on the throat \cite{star3}. 
  
\vspace{-3mm}  
% ===============================================
\section{Conclusion}  
% ===============================================
  
  In our opinion, our joint studies with Alexei Starobinsky, briefly reviewed in the present 
  paper, are sufficiently rich in general observations of common interest. His more specific 
  results concern subtle points of Brans-Dicke solutions (\cite{star3} and Section 4 here) 
  and the existence and nature of singularities outside \wh\ throats in solutions with a 
  phantom Chaplygin gas \cite{star4}, they are also instructive and show Alexei's high 
  level of understanding the physical picture under mathematical complexities.
  
  As to the possible reality of \whs\ in nature, our general impression is that despite the 
  various no-go theorems, this reality is not too strongly suppressed because if throats
  in ghost-free theories are not ruled out, it is an easy task (at least theoretically) to 
  take a certain neighborhood of a throat and to match it with \asflat\ regions of vacuum
  or common ghost-free matter distributions. We can recall, in addition, that there can 
  be dynamic \whs\ whose lifetime is finite but comparable with the age of the Universe
  \cite{BKS23}.
  
  To conclude this paper, we would like to mention some developments that may be said 
  to complement the results described above and (among many others) widen the 
  landscape of \wh\ physics. Thus, Ref.\,\cite{trapped} presented a class of \wh\ models 
  with so-called {\it trapped ghosts\/} where a scalar field with a nonzero potential is 
  phantom in a strong-field region near the throat but is canonical outside it. 
  In \cite{cy1, cy2} it was demonstrated that stationary
  rotating cylindrical \whs\ in GR may be in principle made of phantom-free matter 
  though require rather special conditions for their construction. Lastly, there is a significant
  progress in `\wh\ astronomy', for example, Ref.\,\cite{nov} describes how the sky of
  another universe or a remote part of our own universe could be observed through the 
  throat of a large cosmic \wh.    
    
\vspace{-3mm}  
\subsection*{Acknowledgments}
{Our cordial thanks to Sergei Bolokhov for numerous helpful discussions and remarks.
This work was supported by the RUDN University Project No. FSSF-2023-0003.}
    
\vspace{-3mm}      
% ============================================

\end{document}